# SUBSTRUCTURES AND DENSITY PROFILES OF CLUSTERS IN MODELS OF GALAXY FORMATION

Y.P. Jing[1,2], H.J. Mo[1], G. Börner[1], L.Z. Fang[2]

[1] Max-Planck-Institut für Astrophysik

Karl-Schwarzschild-Strasse 1

85748 Garching, Germany

[2] Department of Physics, University of Arizona

Tucson, AZ 85721, USA

**ABSTRACT**

In this paper we investigate, using high resolution N-body simulations, the density profiles and the morphologies of galaxy clusters in seven models of structure formation. We show that these properties of clusters are closely related to the occurrence of a significant merging event in the recent past. The seven models are: (1) the standard CDM model (SCDM) with $\Omega_0 = 1$, $\Lambda_0 = 0$ and $h = 0.5$; (2) a low-density flat model (FL03) with $\Omega_0 = 0.3$, $\Lambda_0 = 0.7$ and $h = 0.75$; (3) an open model (OP03) with $\Omega_0 = 0.3$, $\Lambda_0 = 0$ and $h = 0.75$; (4) a low-density flat model (FL02) with $\Omega_0 = 0.2$, $\Lambda_0 = 0.8$ and $h = 1$; (5) an open model (OP02) with $\Omega_0 = 0.2$, $\Lambda_0 = 0$ and $h = 1$; (6) a low-density flat model (FL01) with $\Omega_0 = 0.1$ and $\Lambda_0 = 0.9$; (7) an open model (OP01) with $\Omega_0 = 0.1$ and $\Lambda_0 = 0$. We find that the density profiles and morphologies of clusters depend both on $\Omega_0$ and on $\Lambda_0$. For $\Lambda_0 = 0$, these properties are a monotonic function of $\Omega_0$. Clusters in OP01 have the steepest density profiles, their density contours are the roundest and show the smallest center shifts. The other extreme case is SCDM, where clusters show the least steep density profiles and the most elongated contours. For a given $\Omega_0$ ($< 1$), clusters in the flat model (i.e. with $\Lambda_0 = 1 - \Omega_0$) have flatter density profiles and less substructures than in the corresponding open model. Clusters in FL03 have density profiles and center shifts close to those in SCDM, although their density contours are rounder. Our results show that, although cluster density profiles and morphologies are useful cosmological tests, low-density flat models with $\Omega_0 \sim 0.3$, which are currently considered as a successful alternative to SCDM, can produce a substantial fraction of clusters with substructures. This is in contrast to the conception that this kind of models may have serious problem in this aspect.

# 1 INTRODUCTION

Observational studies have demonstrated that a considerable fraction of clusters show evidence of substructures in the galaxy distributions (e.g. Geller & Beers 1982; Dressler & Shectman 1988; West & Bothun 1990) and in the x-ray images (e.g. Jones & Forman 1992, and references therein; Böhringer 1993; Mohr, Fabricant & Geller 1993). The observational data are expected to be improved greatly in the near future. These observations are potentially useful in constraining models of structure formation.

According to current models of structure formation, the mass density of the universe is dominated by a dissipationless component of dark matter. The structure in this component forms by hierarchical gravitational clustering from low amplitude inhomogeneities, with smaller objects collapsing first, and merging to form larger and larger objects. Clusters of galaxies, which are the largest collapsed objects in the universe, are expected to be dynamically young, to show substructures and to have density profiles that differ from those expected from dynamical equilibrium. The deviation from a completely relaxed configuration should, however, be different for different models of structure formation, since the characteristic time of cluster formation depends both on cosmological model and on the power spectrum of initial density fluctuations (White 1992 for a review). Based on the spherical model and the extension of the Press-Schechter formalism, Richstone, Loeb & Turner (1992), Kauffmann & White (1993), and Lacey & Cole (1993) show that the evolution of clusters depends on the cosmological density parameter, with clusters forming earlier in a lower density universe. More realistic modelling has been carried out by numerical simulations. White (1977) and Cavaliere et al. (1986) simulated clusters by mass particles in self-gravitating and initially uniform spheres. West, Oemler & Dekel (1988) simulated cluster evolution in models with various artificial but physically motivated spectra. However, the spatial resolution of their simulation is low. Evrard (1990), Thomas & Couchman (1992) and Katz & White (1993) investigated the evolutions of single clusters, using N-body gas-dynamic simulations. While these simulations account for some physical details of cluster formation, it is at present difficult to obtain a large, statistically fair sample from such a simulation. Based on 8 simulated clusters for each model, Evrard et al. (1993) conclude that the x-ray morphology of clusters depends strongly on the mean mass density of matter in the universe, but depends only weakly on the presence of a cosmological constant. According to their results, the currently popular model with a low density parameter and a nonzero cosmological constant can be ruled out. It is obviously important to carry out such theoretical studies for different models by using large simulations and well defined statistics.

In this paper we investigate, using high resolution N-body simulations, the substructures and density profiles of clusters in seven models of galaxy formation. In Section 2, we describe our simulations and the algorithm for identifying clusters. In Section 3 we show the distribution of mass particles in and around clusters and examine their density profiles, and use different statistical methods to characterize the departure of clusters from a dynamically relaxed configuration. Section 4 gives a brief discussion of our results and summarizes our main conclusions.

## 2 COSMOLOGICAL MODELS AND N-BODY SIMULATIONS

We have run simulations for the Standard CDM model (SCDM) and for six low-density CDM models with or without a cosmological constant. The SCDM model has been popular in the past decade (Davis & Efstathiou 1988) though it seems to have too much small scale clustering power when the density power spectrum is normalized by the $COBE$ observation (Smoot et al. 1991). Low-density models with $\Omega_0 = 0.2 - 0.3$ have been considered as alternative models which seem to be compatible with most observational data on galaxy clustering and on the cosmic microwave background anisotropy (e.g., Bahcall 1994; Wright et al. 1992; Efstathiou et al. 1992). However, since structures form earlier in a universe with lower $\Omega_0$, it is unclear whether or not clusters in such models show too little substructures to be compatible with the observation. The six low-density models considered in present paper have $\Omega_0 = 0.3, 0.2, 0.1$, with $\Lambda_0 = 0$ or $\Lambda_0 = 1 - \Omega_0$. In the following, they will be referred as FL03 (FLat model with $\Omega_0 = 0.3$), OP03 (OPen model with $\Omega_0 = 0.3$), FL02, OP02, FL01 and OP01. The last two models are not physically motivated; they are included for comparison. In Table 1, we list the values of $\Omega_0$, $\Lambda_0$ and $h$† for each model. For all models except FL01 and OP01, we use the transfer functions given by Bardeen et al. (1986) for the corresponding model parameters. The primordial power spectrum is assumed to have the Harrison-Zel'dovich form. For FL01 and OP01, the linear power spectra given by their own transfer functions do not have the correct shape to match the observation of galaxy clustering. In these two cases, the initial power spectra are taken to be the same as that for SCDM. The initial spectra for all cases are normalized, so that $\sigma_8$ (the $rms$ density fluctuation in a sphere of radius $8\,h^{-1}$Mpc linearly evolved to the present time) has the values listed in Table 1. The normalization for the SCDM model is consistent with the mass measurements of rich clusters (White et al. 1993). This normalization is, however, not compatible with the $COBE$ observation. When normalized by $COBE$ observation, this model gives too many rich clusters. The two flat

---

† The Hubble constant $H_0$ is written as $H_0 = 100\,h\,\mathrm{km\,s^{-1}Mpc^{-1}}$

models with $\Omega_0 = 0.3$ and 0.2 so normalized are within the $1\sigma$ uncertainty level of the *COBE* measurement, and is compatible with most observational data.

The simulations were performed by using a P$^3$M code. The code was designed according to the standard method described by Hockney & Eastwood (1981) and Efstathiou et al. (1985). The details of the code can be found in Jing & Fang (1994; hereafter JF94). Here we describe only several important parameters for our present simulations.

The simulations were done in a cubic box of size $128\,h^{-1}$Mpc, with force resolution $\eta = 0.1\,h^{-1}$Mpc. We have run 3 realizations for the SCDM model. In the first two runs, $100^3$ particles were used. In the third run, we use $128^3$ particles, to see possible effect of finite particle number on our statistics. In fact, we found no statistical difference between the first two and the last runs. Therefore, in the following we will not discuss them separately. For the six low-$\Omega_0$ models, we use $64^3$ particles in each simulation and make five runs for each model. Since the mean mass densities in the low-density models are lower than that in the SCDM model, the number of particles in clusters of similar physical properties (e.g., mass/velocity dispersion) is higher in the low-density models if we use the same number of particles in the simulations. This is why we use less particles in the low-density models than in the SCDM model. The simulation parameters for each model are given in Table 1.

Our simulations are suitable for the study of the internal properties of clusters. The formation and evolution of clusters is sensitive only to the linear density perturbations on scales of about $10\,h^{-1}$Mpc. The initial conditions in our simulations can correctly describe the linear density perturbation over a wide range of scales around the cluster scale. The P$^3$M code with its high force resolution is able to follow accurately the non-linear dynamics of clusters around their cores. In our simulations, each rich cluster contains typically about 700 or more particles within the Abell radius, so that the density field of a cluster is well sampled.

To identify cluster-like dark halos, we use the same procedure as described in JF94. The procedure is divided into two steps: we first find groups based on the *friends-of-friends* algorithm and then search for the gravitational potential minima around these groups. The details of this procedure are given in JF94.

In the *friends-of-friends* algorithm, the link parameter $b$ is chosen to be 0.2 times the mean particle separation. As shown in JF94, the groups identified in this way can have irregular shapes. Some of the groups may have their centers of mass not located in dense regions. This can happen, for example, when a group consists of two dark halos connected by a thin bridge. Since an accurate determination of the centers of dark halos is important in our analysis, we need to treat this kind of situation.

Since massive halos are usually located at the gravitational potential minima, we can hope to find these halos by searching for the potential minima around the groups already identified. We place a cube of side $12.8\,h^{-1}$Mpc and a grid of $64^3$ uniform meshes around each group. The mass density field will be sampled on the meshes. We smooth the original particle distribution by a Gaussian kernel

$$W(r,s) = \frac{1}{(2\pi)^{3/2}s^3}\exp\left(-\frac{r^2}{2s^2}\right), \qquad (1)$$

where $s$ is the smoothing length. To have a smooth density field with the substructures reasonably resolved, we use a spatially varying smoothing length. For a particle $i$, the smoothing length $s_i$ is chosen to be the local mean separation $d_i$ of the five nearest neighbours of the particle. Because the density field is sampled on meshes, we require that $s_i$ be larger than the cell size. The density on an arbitrary mesh $j$ is then given by

$$\rho_j = \sum_i m_i W(r_{ji}, s_i), \qquad (2)$$

where $r_{ij}$ is the separation between cell $j$ and particle $i$ of mass $m_i$. The gravitational potential on the grids is obtained, as in the N-body simulation, by the FFT technique. To eliminate the boundary effects, we consider only the central cubic volume of side length $7.7\,h^{-1}$Mpc, which is about 5 times the Abell radius. A cell is identified as a potential minimum if its potential value is smaller than those of its 26 neighbors. The coordinates of the potential minima are then the positions of the dark halos we have identified. We use the 3-D velocity dispersion $\sigma_v$, defined as the velocity dispersion of particles within a radius of $0.5\,h^{-1}$Mpc, to characterize a dark halo. We delete those halos which are within a radius of $1\,h^{-1}$Mpc from a larger halo (with larger $\sigma_v$). These small halos are regarded as substructures of the larger halo. We construct our dark halo samples which are complete to a given value of $\sigma_v$. Since an accurate determination of halo centers is important for our following statistical analyses, we repeat the minima searching around the halos using a cube of $6.4\,h^{-1}$Mpc and $128^3$ meshes. The final accuracy of the center positions is better than $0.05\,h^{-1}$Mpc.

## 3 THE DENSITY PROFILE AND MORPHOLOGY OF CLUSTERS

As an illustration of cluster formation histories, we show in Figure 1 the particle distributions around the 4 biggest clusters in one realization for models SCDM, FL03 and OP01. The phases of the initial density perturbations are the same for the three models. We note that the clusters in the three models do not have one-to-one correspondence in their *Lagrangian* spatial regions, because the merging histories of clusters are different. For example, one cluster in OP01 may contain several clusters in SCDM. Particle distributions are shown for four different redshifts, as shown in the figures. For comparison, we draw only part of the particles randomly selected from the simulated clusters in SCDM and OP01, so that each panel for these two models contains the same number of particles as the corresponding panel for FL03. The clusters in OP01 appear to be compact and round, while those in the other two models appear to be less concentrated and more irregular. From the evolution sequences, we see that clusters in OP01 tend to be dynamically old, with negligible merging of clumps since quite a high redshift. In SCDM, however, many clusters are still in their early stages of formation, with a lot of clumps falling in during the quite recent past. The merging in FL03 seems to be less frequent than that in SCDM. However, recent merging of clumps does happen for a substantial fraction of the simulated clusters in this model. The properties of clusters in the other four models are intermediate between the FL03 and OP01 models. In the following subsections, we will use different methods to quantify these differences.

## 3.1 The density profiles of clusters

To quantify the density profiles of clusters, we measure the cross-correlation function between cluster and mass particles. For each cluster, we count the number of particles $N(r)$ within two spherical shells of radii $r - \Delta r/2$ and $r + \Delta r/2$ around the cluster center. The cross correlation $\xi_{\rm cm}$ at radius $r$ is defined as

$$\xi_{\rm cm}(r) = \frac{N(r)}{N_{\rm exp}(r)} - 1, \qquad (3)$$

where $N_{\rm exp}(r)$ is the expected number of particles within these two shells, calculated from the mean number density of particles in the simulation. Figure 2 shows the cross-correlation functions for clusters within six bins of different velocity-dispersion. The amplitude of the cross-correlation depends both on the velocity dispersion of clusters and on cosmological model. The former dependence is due to biasing, while the latter is due to the fact that, for a given velocity dispersion, the mass of a cluster is approximately the same in different models and the correlation is stronger for a model with lower $\Omega_0$. However, for a given model, the shape of the density profiles does not depend strongly on the velocity dispersions of clusters. We have fitted the cross-correlation function by a power law $\xi_{\rm cm} \propto r^{-\beta}$ in the range $r = 0.2$ to $1.0 \, h^{-1}$Mpc. The means and the standard deviations of the means of the slopes $\beta$ among the profiles shown in Fig. 2 are listed in the first line of Table 2. Since the density profiles are not a pure power law, we have also made a similar fitting for data points in the range $50 \leq \xi_{cm} \leq 3000$. The results are listed in the second line of Table 2. It is clear from Table 2 and Fig.2 that, for $\Lambda_0 = 0$, a model with lower $\Omega_0$ has systematically steeper density profile. For a given $\Omega_0$, clusters in a flat model have flatter density profiles. The difference in slope between SCDM and FL03 is only marginal when the fit is made for a given range of $r$. It becomes significant when the fit is made for a given range of $\xi_{cm}$. It should be pointed out that in real observation the slope can be measured only for a given range of $r$, if the mean density of the universe is unknown. It is interesting to note that the mean slope for FL03 is closer to SCDM than to OP03.

## 3.2 The shapes of dark halos

Assuming that the isodensity surface of a cluster is approximately described by a triaxial ellipsoid of the form:

$$\frac{x_1^2}{a_1^2} + \frac{x_2^2}{a_2^2} + \frac{x_3^2}{a_3^2} = 1; \qquad a_1 \leq a_2 \leq a_3, \tag{4}$$

where $a_1$, $a_2$, and $a_3$ are the lengths of the three axes, we can examine the axial ratios $a_1/a_3$ and $a_2/a_3$ for the cluster. We use these axial ratios as a measure of cluster shapes. The principal axes $x_1, x_2$ and $x_3$ are generally unknown *a priori* in the simulations. These axes, as well as the axial ratios can be obtained by finding the eigenvectors and eigenvalues of a matrix $\{M_{\alpha\beta}\}$:

$$M_{\alpha\beta} = \sum_{i,j=1}^{N_c} X_\alpha^i X_\beta^j; \qquad \alpha, \beta = 1, 2, 3, \tag{6}$$

where $X_\alpha^i$ is the $\alpha$-axis coordinate of the $i$th particle, relative to the center of mass. One difficulty in getting the axial ratios is how to select the particles $1, 2, ..., N_c$, for we do not know the isodensity surfaces of the ellipsoid. Here we use an iteration method similar to that in Katz (1991). First we take all particles around a cluster center within the virial radius $r_v$ interior to which the mean overdensity of particles is $176\Omega_0^{-0.6}$. We calculate the principal axes $\{x_\alpha\}$ and the axial ratios $a_1/a_3$ and $a_2/a_3$ for these particles by solving the eigen equation of $\{M_{\alpha\beta}\}$. Then we calculate the new principal axes and the axial ratios using particles within an ellipsoid (Equation 4) with the principal axes and the axial ratios just determined (we set $a_3 = r_v$). We repeat the same calculation for the updated ellipsoid, until the axial ratios $a_1/a_3$ and $a_2/a_3$ converge to an accuracy of 1%.

Figure 3 shows the distributions of the axial ratios for the 50 clusters of the highest velocity dispersions in each realization of the seven models. The mean values of $(a_1/a_3, a_2/a_3)$

are (0.44, 0.60) for SCDM, (0.47, 0.65) for FL03, (0.51, 0.67) for OP03, (0.51, 0.67) for FL02, (0.53, 0.70) for OP02, (0.53, 0.68) for FL01, and (0.57, 0.73) for OP01. To have a quantitative test for the differences among these distributions, we calculate, using the Kolmogorov-Smirnov (K-S) test, the probability $P(K-S)$ that two models, say SCDM and LCDM, have the same $a_1/a_3$ (or $a_2/a_3$) distribution. The probabilities are given in Table 3, with those listed above the diagonal for the $a_1/a_3$ distribution and those listed below for the $a_2/a_3$ distribution. It is clear that the clusters in OP01 have systematically higher axial ratios than those in the other six models. This means that on average the clusters in this model are the roundest. The clusters in SCDM are the most elongated. The axial-ratio distributions for FL03 are most close to those for SCDM. The other sample pairs, which have similar axial-ratio distributions, are (OP03, FL02), (OP03, OP02), (FL02, FL01), and (OP02, FL01). It is interesting to note that among these sample pairs, there is no pair in which the two models have the same $\Omega_0$. Clusters in a flat model appear to be significantly more elongated than in the corresponding open model. Among open (or flat) models, although some neighboring models (e.g. OP03 and OP02) have statistically compatible axial-ratio distributions, clusters in a lower $\Omega_0$ model appear to be rounder.

## 3.3 Cluster shapes in projection

In this subsection we study further the morphology of clusters from their projected density distribution. As we will see clearly later, the projected density distribution of clusters is closely related to the 3-D cluster shapes discussed in Section 3.2 and to the X-ray images to be discussed in Section 3.4.

We first obtain a smoothed 3-D density distribution in a cube of side $3.2\,h^{-1}$Mpc according to the techniques given in Section 2. Here we use a smoothing length $s_i$ which is twice the local mean separation of particles $d_i$ (see Section 2). The smoothing length is chosen after several trials, so that the resulting density field is a compromise between good spatial resolution and sufficient smoothness. Our results remain nearly unchanged if we change the value of $s_i$ by a factor of two. We then calculate the square of the density in each cell and project the density square on a given plane. The reason for using density square in the projection is to have a more direct comparison with the x-ray surface density (the x-ray luminosity in a cell is proportional to the square of the gas density in the cell). Very similar results were obtained when we used density itself in the projection. For brevity, we will denote the projected field of density square by $\mathcal{DS}$.

Figure 4 shows the contours of the $\mathcal{DS}$ distributions for the 12 clusters at $z=0$

shown in Figure 1. The projection is made from the 3-D smoothed density square in the $3.2\,h^{-1}$Mpc box. The projection surface is the same as that in Figure 1, so that we can compare these two figures directly. The most noticeable point is that the $\mathcal{DS}$ contours reflect clearly recent merging events. For example, as we can see from Figure 1, each of the three clusters, cl1, cl3 in FL03 and cl4 in SCDM, has a significant merging event in the recent past. Their $\mathcal{DS}$ contours all show peanut-like shapes near the centers and have significant center shifts. In contrast, clusters without significant recent mergers, like those in OP01, show round contours with small center shifts. Small merging events, like those in cl1 and cl2 in the SCDM model, and cl2 in the FL03 model, also show up as distortions in the $\mathcal{DS}$ contours. From the above results, we see clearly that the shape and the center shift of the $\mathcal{DS}$ contours can be used as an indicator of the dynamical state of clusters. We have inspected the $\mathcal{DS}$ contours for all clusters. Many clusters in the SCDM model show elongated contours with large center shifts. The contours of most clusters in the OP01 model are round with small center shifts. The clusters in FL03 tend to be rounder than those in SCDM, but much more irregular than thoe in OP01. About 30 percent of clusters in FL03 have distorted contours that look like cl1 and cl4 in SCDM, and cl1 and cl3 in the FL03, shown in Fig. 4. The corresponding percentages are about 50 percent for SCDM and less than 10 percent for OP01. The fraction of substructures in the other models is between 10 and 30 percent.

For each cluster, we consider 9 contours of equal $\mathcal{DS}$ values. We first put pixels in a successively decreasing order of $\mathcal{DS}$ values. The $i$th contour level is chosen to be equal to the $\mathcal{DS}$ value of the pixel with order number $[6+2(i-1)]^2$. The last contour corresponds to a linear scale of about $1\,h^{-1}$Mpc. The center of each contour and its minor and major axes are determined from the moments with respect to the x- and y- axes. The axis ratio (defined as the ratio of the minor axis to the major axis) for each cluster is a weighted average of the axis ratios of the 8 inner contours. The weight of each contour $w_i$ is proportional to the sum of the $\mathcal{DS}$ values of the pixels between this contour and the adjacent outer contour. The axis shift for each cluster is defined as $\sum_{i=1}^{8} w_i[(x_i - \bar{x})^2 + (y_i - \bar{y})^2]$, where $(x_i, y_i)$ is the center of the $i$th contour, $\bar{x} = \sum w_i x_i$, $\bar{y} = \sum w_i y_i$. In Figure 5 we show the histograms for the axis ratios and center shifts for the 50 clusters of the highest velocity dispersions in each realization of the seven models. Here all pixels inside the $3.2\,h^{-1}$Mpc $\times$ $3.2\,h^{-1}$Mpc square that satisfy the criterion on $\mathcal{DS}$ are used. In this case, the center shifts for some clusters are quite large, because these clusters contain two or more separate components.

From the figure, it is clearly seen that clusters in models with lower $\Omega_0$ tend to be rounder and to have smaller center shifts. The effect of the cosmological constant is also seen in a comparison between a flat model and the corresponding open model. The clusters in the SCDM model tend to be the most elongated, with the largest center shifts. An inspection of the $\mathcal{DS}$ contours showed that the tail of large center shift in OP01 and OP02 is due to clusters with separate components at large separations. We have done a K-S test for the distributions shown in Fig. 5. The results are given in Table 4. The items listed above the diagonal are for axis ratio and those below for center shift. The total number of data points used in the K-S test is quite large for each case, so the differences among the seven models are all significant (at 95% level). However, the axis-ratio distributions are the most similar $[P(K-S) > 0.01]$ among the flat models FL03, FL02 and FL01, between the open models OP03 and OP02, and between FL02 and OP03. The center-shift distributions are most similar in four pairs of models: (FL03, FL02), (OP03, OP02), (FL02, OP03) and (FL03, OP03). The similarity of the center-shift distribution between FL03 and OP03 indicates that the effect of $\Lambda_0$ on substructure is smaller in a higher $\Omega_0$ universe, as expected. These dependence of $\mathcal{DS}$ contours on $\Omega_0$ and $\Lambda_0$ are in good agreement with what we found in the last two subsections.

In Figure 6, we show the same histograms as in Fig. 5, but here we only consider pixels inside a circle (centered on the cluster center) with a radius of $0.5\,h^{-1}$Mpc. This case is considered, because we want to examine the density field in the central region of a cluster. The results of K-S test for this case are presented in Table 5. It is clearly seen that the dependence on models is similar to that shown in Figure 5, except that the center shifts are systematically smaller in all models. This occurs because in this case only a smaller and perhaps more relaxed region is considered for each cluster. The tail of large center shift in OP01 and OP02 shown in Fig. 5 disappears because some components at large separations are removed. In this case the center-shift distributions for SCDM and FL03 become similar.

### 3.4 X-ray images using hydrostatic models

We construct x-ray images for a cluster under the following assumptions: 1) the intracluster gas is in hydrostatic equilibrium under the gravitational potential of dark matter; 2) the gas is polytropic with an index $\gamma$; 3) the central gas temperature is equal to the virial temperature; and 4) the gas composition is primordial. The formulae for such a calculation can be found in Cavaliere et al. (1986). We point out that a quantitative comparison between theoretical predictions and x-ray observations (e.g., Jones & Forman 1992) can be made only when these assumptions are justified. As shown below, the x-ray images so obtained reflect clearly the recent evolutionary history of clusters.

In the construction of an x-ray image, one crucial step is to calculate the gravitational potential field around a cluster. Here we use a procedure exactly the same as described in Section 2, except that here we consider a box of $6.4\,h^{-1}$Mpc around a cluster and use $128^3$ meshes for the potential field. We take the polytropic index $\gamma = 1$, namely, an isothermal gas distribution. The calculation of the x-ray luminosity in each cell is then straightforward. We will consider only the central cube of $3.2\,h^{-1}$Mpc in our following discussion, in order to reduce the boundary effect on the calculation of the potential.

The projection of the x-ray luminosity is done in the same way as that of the density field (see Section 3.3). We show, in Figure 7, the x-ray iso-intensity contours for the same 12 clusters as shown in Figure 4. The panels in the two figures are arranged in the same way. The most interesting point shown in Fig. 7 is that all important features seen in the $\mathcal{DS}$ contours (Fig. 4) are also seen in the x-ray contours. For examples, cl1 and cl3 in FL03, and cl4 in SCDM also show peanut-like shapes and large center shifts in the x-ray contours. The clusters in OP01 still have round and compact shapes. The only difference is that the x-ray contours are rounder than the $\mathcal{DS}$ contours. This is expected, since the spatial distribution of the gas should be rounder than that of dark matter.

In Figure 8 we show the distributions of the axis ratios and center shifts of the x-ray contours. The histograms are obtained in the same way as described in Section 3.3. The analysis of the contours is similar to that for Fig. 5. Here the weight given to each contour is proportional to the total x-ray luminosity between this contour and the adjacent outer contour. Since the x-ray emission is mainly concentrated in the central region of a cluster, our results will not change significantly, if we do an analysis similar to that used for Figure 6. From the figure, and the corresponding K-S test presented in Table 6, we see that the distributions of axis ratios and center shifts in the x-ray contours have the same model dependence as those in the $\mathcal{DS}$ contours shown in Figure 6. We see again that the center-shift distribution of FL03 is similar to SCDM. On the same scale, the x-ray contours are rounder and have smaller center shifts than the corresponding $\mathcal{DS}$ contours.

## 4. DISCUSSION AND CONCLUSIONS

Based on 8 clusters from N-body gasdynamic simulations, Evrard et al. (1993) found that the x-ray images of clusters in low-density models (with $\Omega_0 = 0.2$) are much more regular, spherically symmetric and centrally condensed than those in an Einstein-de Sitter model ($\Omega_0 = 1$), with only a weak dependence on a possible cosmological constant. They used this result to argue against the low-density models even with a cosmological constant.

The general dependence of substructures on cosmological parameters found in our work is in agreement with that they found. However, with our larger samples, we also found, in contrast with their claims, that the clusters in the currently interesting low-$\Omega_0$ models with $\Lambda_0$ (e.g. FL03), have density profiles and center shifts of contours that are similar to those in the SCDM model, although the density and x-ray image contours are indeed rounder. In particular, we do see a large fraction of clusters in the flat models with $\Omega_0 = 0.3$ showing significant substructures. Our results show that it is still premature to claim that this kind of models may have serious problem in this aspect. It is still unclear to us what causes this discrepancy. The apparent differences between the two studies are: (1) They have assumed a high fraction (50 percent) of mass in baryons for their low-density models, which might have the effect of making the x-ray images rounder. (2) Since their simulation boxes are small, the periodic boundary condition used in their code may reduce the clustering power on large scales. This effect might be more severe in the low-density models. (3) The number of clusters in their work is small. (4) In our work we have based our analysis on the dark-matter distribution. Some of the discrepancy may be explained, if the x-ray emission does not trace well the potential of dark matter. Some of our assumptions in constructing the x-ray images may not be valid, and the polytropic model may have erased some small components. We expect that the real x-ray images will have properties between the $\mathcal{DS}$ and our x-ray images.

In a recent work by Crone, Evrard & Richstone (1994) which analyzed the density profiles of clusters in different cosmological models with various power-law power spectra, it is also found that the density profiles of clusters in a flat low-density model are not very different from those in an Einstein-de Sitter model. Our results agree with theirs quantitatively when we fit the density profiles by a power law in the same density contrast range. As we have shown in subsection 3.1, the differences in slopes among models become smaller if we fit the data in a given range of radii, because the density profile is steeper at larger radius.

In this paper, we have only considered the properties of clusters in theoretical models. A comparison of our model predictions with observations will be given in a future work. It is perhaps necessary to point out that it is not straightforward to compare simulation results with observations presently available. Although x-ray observations of clusters have revealed that a large fraction of clusters show substructures in their x-ray images (e.g. Jones & Forman 1992; Böhringer 1993), we still lack a quantitative study of substructures for a large and statistically complete sample. Furthermore, a large fraction of the substructures found by Jones & Forman are on small scales, with typical projected separations less than about $0.3\,h^{-1}$Mpc (C.S. Grant & C. Jones, private communication). On these scales, current numerical simulations are still not able to treat the relevant physical processes realistically. For example, the core radii (typically $0.2\,h^{-1}$Mpc) of the x-ray emission observed in real clusters are still not produced by current N-body hydrodynamic simulations (Evrard 1990; Thomas & Couchman 1992; Katz & White 1993). The density profiles are closely related to the cluster-galaxy cross-correlation function $\xi_{cg}$ (Seldner & Peebles 1977). However, observational results of $\xi_{cg}$ are still uncertain on cluster scales (Lilje & Efstathiou 1988). The comparison between models and the observations is further complicated by the fact that galaxies may not trace the mass distribution in a simple way. With our present knowledge of galaxy formation, it is still unclear how to identify galaxies reliably from numerical simulations. The situation will be improved in the near future with more data available from x-ray observations and from weak gravitational lensing (e.g. Kaiser & Squires 1993).

In summary we have investigated, using high resolution N-body simulations, the density profiles and the shapes of clusters to show their dependence on models of structure formation. We have shown that these properties of clusters are closely related to the occurrence of significant merging events to a cluster in the recent past. We found that for $\Lambda_0 = 0$, clusters in a lower $\Omega_0$ universe show steeper density profiles, rounder morphologies and larger center-shifts in their density contours. Clusters in a low-density flat universe have flatter density profiles, less round morphologies and larger center shifts than those in the corresponding open universe. In particular, we found that clusters in a flat low density model with $\Omega_0 \sim 0.3$ (and with cosmological constant) have density profiles and center shifts (in their density contours) similar to those in an Einstein-de Sitter model, although the cluster shapes are rounder. Our results show that, although the density profiles and morphologies of clusters depend on models of structure formation, the flat low density model with $\Omega_0 \sim 0.3$, which is currently considered as a successful alternative to the standard CDM model, can produce a substantial fraction of clusters with substructures. Given current observational situation and the fact that we do see a considerable fraction of clusters showing substructures in such a model, it is still premature to claim that this kind of models has serious problem in this aspect, in contrast to current belief and to some previous results.

## Acknowledgements

We would like to thank Neta Bahcall, Martin Haehnelt, especially Simon White for helpful discussions, Carolyn Grant and Christine Jones for communicating their statistical results of substructures in advance of publication. YPJ acknowledges the receipt of an Alexander-von-Humboldt research fellowship. He also thanks the World Laboratory for a scholarship during early stages of this work.

# References

Bahcall N.A., 1994, in Gottlöber S., Mücket J., eds., Large Scale Structure in the Universe, Proceeding of the 11th Potsdam Cosmology Workshop. to be published

Bardeen J., Bond J.R., Kaiser N., Szalay A.S., 1986, ApJ, 304, 15

Böhringer H., 1993, in Silk J., Vittorio N., eds., Galaxy Formation, Proceeding of the E. Fermi Summer School

Cavaliere A., Santangelo P., Tarquini G., Vittorio N., 1986, ApJ, 305, 651

Crone M.M., Evrard A.E., Richstone D.O., 1994, ApJ, in press

Davis M., Efstathiou G., 1988, in Rubin V.C., Coyne G.V., eds., Large Scale Motions in the Universe. Princeton Univ. Press, Princeton, p.439

Dressler A., Schectman S., 1988, AJ, 95, 985

Efstathiou G., Bond J.R., White S.D.M., 1992, MNRAS, 258, P1

Efstathiou G., Davis M., Frenk C.S., White S.D.M., 1985, ApJS, 57, 241

Evrard A.E., 1990, ApJ, 363, 349

Evrard A.E., Mohr J.J., Fabricant D.J., Geller M.J., 1993, ApJ, 419, L9

Geller M.J., Beers T.C., 1982, PASP, 94, 421

Hockney R.W., Eastwood J.W., 1981, Computer Simulations Using Particles. McGraw-Hill Inc, New York

Jing Y.P., Fang L.Z., 1994, ApJ, 432, 438 (JF94)

Jones C., Forman W., 1992, in Fabian A.C., ed., Clusters and Superclusters of Galaxies. Kluwer, Dordrecht, p.49

Kaiser N., Squires G., 1993, ApJ, 404, 441

Katz N., 1991, ApJ, 368, 325

Katz N., White S.D.M., 1993, ApJ, 412, 455

Kauffmann G., White S.M.D., 1993, MNRAS, 261, 921

Lacey, C., Cole, S. 1993, MNRAS, 262, 627

Lilje P.B., Efstathiou G., 1988, MNRAS, 231, 635

Mohr, J.J., Fabricant, D.G., Geller, M., 1993, ApJ, 413, 492

Richstone D., Loeb A., Turner E.L., 1992, ApJ, 392, 477

Seldner M., Peebles P.J.E., 1977, ApJ, 215, 703

Smoot G.F., et al., 1992, ApJ, 396, L1

Thomas P.A., Couchman H.M.P, 1992, MNRAS, 257, 11

West M., Bothun G., 1990, ApJ, 350, 36

West M., Oemler A., Dekel A., 1988, ApJ, 327, 1

Wright E.L., et al., 1992, ApJ, 396, L13

White, S.D.M., 1977, MNRAS, 177, 717

White, S.D.M., 1992, in Fabian A.C., ed., Clusters and Superclusters of Galaxies. Kluwer, Dordrecht, p.1

White, S.D.M., Efstathiou, G., Frenk, C.S. 1993, MNRAS, 262, 1023

# Figure captions

**Figure 1 (a–c).** The evolution of particle distributions around the four most massive clusters in one realization (with the same random phases) of each of the three models. The epochs at output are at redshifts $z = 0.5, 0.33, 0.13, 0$. The scales are the physical coordinates in units of $h^{-1}$Mpc.

**Figure 2.** The cross-correlation function $\xi_{\rm cm}(r)$ between clusters and mass, which serves as a measure of the mean density profile of clusters. 48 most massive clusters in each realization are considered, and these clusters are divided into 6 groups in the increasing order of the velocity dispersion. Panels (a) to (f) show $\xi_{\rm cm}(r)$ for clusters from small to large velocity dispersion.

**Figure 3.** The distributions of the axial ratios $a_1/a_3$ and $a_2/a_3$ of the virialized dark matter halos of the 50 clusters in each realization with the highest velocity dispersions.

**Figure 4.** The projected density-square contours for the twelve clusters at $z = 0$ shown in Figure 1. The $i$-th contour level is chosen to be $10^{-0.1 \times i}$ times the maximum $\mathcal{DS}$ value. Scale units are in $h^{-1}$Mpc.

**Figure 5.** The distributions of the mean axis ratios and center shifts of the projected isodensity contours in a central square region of $3.2\, h^{-1}$Mpc $\times\, 3.2\, h^{-1}$Mpc around clusters.

**Figure 6.** The same as Figure 5, except that the calculation is done in a central circular region with radius $0.5\, h^{-1}$Mpc around clusters.

**Figure 7.** The x-ray iso-intensity contours of the twelve clusters shown in Fig. 4. The contour levels are chosen in the same way as in Fig. 4. Scale units are in $h^{-1}$Mpc.

**Figure 8.** The distributions of mean axis ratios and center shifts of the x-ray iso-intensity contours.

Table 1. Parameters of the simulations

| | SCDM | FL03 | OP03 | FL02 | OP02 | FL01 | OP01 |
|---|---|---|---|---|---|---|---|
| Box size | $128\,h^{-1}$Mpc | $128\,h^{-1}$Mpc | $128\,h^{-1}$Mpc | $128\,h^{-1}$Mpc | $128\,h^{-1}$Mpc | $128\,h^{-1}$Mpc | $128\,h^{-1}$Mpc |
| No. of particles | $100^3(128^3)$ | $64^3$ | $64^3$ | $64^3$ | $64^3$ | $64^3$ | $64^3$ |
| No. of meshes | $256^3$ | $128^3$ | $128^3$ | $128^3$ | $128^3$ | $128^3$ | $128^3$ |
| Force Resolution | $0.1\,h^{-1}$Mpc | $0.1\,h^{-1}$Mpc | $0.1\,h^{-1}$Mpc | $0.1\,h^{-1}$Mpc | $0.1\,h^{-1}$Mpc | $0.1\,h^{-1}$Mpc | $0.1\,h^{-1}$Mpc |
| No. of realizations | 2 (1) | 5 | 5 | 5 | 5 | 5 | 5 |
| Initial redshift | 8 | 8 | 13 | 11.7 | 23.4 | 26 | 76.7 |
| Time steps | 400 | 400 | 520 | 585 | 585 | 520 | 767 |
| $\Omega_0$ | 1 | 0.3 | 0.3 | 0.2 | 0.2 | 0.1 | 0.1 |
| $\Lambda_0$ | 0 | 0.7 | 0 | 0.8 | 0 | 0.9 | 0 |
| $h$ | 0.5 | 0.75 | 0.75 | 1 | 1 | — | — |
| $P_i(k)$ | SCDM | FL03 | OP03 | FL02 | OP02 | SCDM | SCDM |
| Normalization $\sigma_8$ | 0.6 | 1 | 1 | 1 | 1 | 1 | 1 |

Table 2. The slope of density profile

| | SCDM | FL03 | OP03 | FL02 | OP02 | FL01 | OP01 |
|---|---|---|---|---|---|---|---|
| $0.2 < r < 1.0\,h^{-1}$Mpc | $2.32\pm 0.04$ | $2.37\pm 0.05$ | $2.52\pm 0.03$ | $2.41\pm 0.04$ | $2.61\pm 0.04$ | $2.63\pm 0.04$ | $2.92\pm 0.03$ |
| $50 < \xi_{cm} < 3000$ | $2.34\pm 0.01$ | $2.51\pm 0.03$ | $2.68\pm 0.04$ | $2.57\pm 0.02$ | $2.77\pm 0.03$ | $2.84\pm 0.02$ | $3.06\pm 0.03$ |

Table 3. K-S test of the halo shapes

| | SCDM | FL03 | OP03 | FL02 | OP02 | FL01 | OP01 |
|---|---|---|---|---|---|---|---|
| SCDM | — | $0.40\times 10^{-2}$ | $0.54\times 10^{-8}$ | $0.26\times 10^{-6}$ | $0.25\times 10^{-11}$ | $0.52\times 10^{-9}$ | $0.31\times 10^{-20}$ |
| FL03 | 0.11 | — | $0.10\times 10^{-2}$ | $0.70\times 10^{-2}$ | $0.13\times 10^{-5}$ | $0.57\times 10^{-4}$ | $0.45\times 10^{-12}$ |
| OP03 | $0.47\times 10^{-2}$ | 0.49 | — | 0.27 | 0.23 | 0.34 | $0.21\times 10^{-5}$ |
| FL02 | $0.99\times 10^{-2}$ | 0.23 | 0.62 | — | $0.39\times 10^{-1}$ | 0.20 | $0.31\times 10^{-7}$ |
| OP02 | $0.20\times 10^{-5}$ | $0.26\times 10^{-2}$ | $0.50\times 10^{-1}$ | $0.60\times 10^{-1}$ | — | 0.39 | $0.73\times 10^{-4}$ |
| FL01 | $0.27\times 10^{-2}$ | $0.37\times 10^{-1}$ | 0.46 | 0.87 | $0.34\times 10^{-1}$ | — | $0.82\times 10^{-4}$ |
| OP01 | $0.42\times 10^{-10}$ | $0.31\times 10^{-7}$ | $0.21\times 10^{-4}$ | $0.12\times 10^{-5}$ | $0.18\times 10^{-1}$ | $0.40\times 10^{-5}$ | — |

Table 4. K-S test of the density-sqaure contours

|      | SCDM | FL03 | OP03 | FL02 | OP02 | FL01 | OP01 |
|------|------|------|------|------|------|------|------|
| SCDM | —    | 0    | 0    | 0    | 0    | 0    | 0    |
| FL03 | $0.50 \times 10^{-7}$ | — | $0.13 \times 10^{-2}$ | $0.14 \times 10^{-1}$ | $0.15 \times 10^{-6}$ | $0.31 \times 10^{-1}$ | $0.91 \times 10^{-12}$ |
| OP03 | $0.22 \times 10^{-7}$ | $0.19$ | — | $0.35 \times 10^{-1}$ | $0.13$ | $0.32 \times 10^{-4}$ | $0.25 \times 10^{-4}$ |
| FL02 | $.80 \times 10^{-11}$ | $0.81 \times 10^{-1}$ | $0.12$ | — | $0.42 \times 10^{-2}$ | $0.10 \times 10^{-1}$ | $0.15 \times 10^{-6}$ |
| OP02 | $0.50 \times 10^{-12}$ | $0.41 \times 10^{-4}$ | $0.31 \times 10^{-1}$ | $0.13 \times 10^{-2}$ | — | $0.71 \times 10^{-5}$ | $0.60 \times 10^{-2}$ |
| FL01 | 0 | $0.11 \times 10^{-9}$ | $0.15 \times 10^{-6}$ | $0.15 \times 10^{-6}$ | $0.50 \times 10^{-2}$ | — | $0.84 \times 10^{-13}$ |
| OP01 | 0 | 0 | $0.16 \times 10^{-13}$ | 0 | $0.44 \times 10^{-7}$ | $0.11 \times 10^{-5}$ | — |

Table 5. K-S test of the density-sqaure contours within 0.5 $h^{-1}$Mpc

|      | SCDM | FL03 | OP03 | FL02 | OP02 | FL01 | OP01 |
|------|------|------|------|------|------|------|------|
| SCDM | —    | 0    | 0    | 0    | 0    | 0    | 0    |
| FL03 | $0.13$ | — | $0.11 \times 10^{-2}$ | $0.10 \times 10^{-1}$ | $0.20 \times 10^{-6}$ | $0.26 \times 10^{-1}$ | $0.19 \times 10^{-12}$ |
| OP03 | $0.57 \times 10^{-3}$ | $0.26 \times 10^{-1}$ | — | $0.81 \times 10^{-1}$ | $0.47 \times 10^{-1}$ | $0.25 \times 10^{-4}$ | $0.25 \times 10^{-4}$ |
| FL02 | $0.42 \times 10^{-6}$ | $0.11 \times 10^{-2}$ | $0.11$ | — | $0.16 \times 10^{-2}$ | $0.60 \times 10^{-3}$ | $0.60 \times 10^{-7}$ |
| OP02 | $0.16 \times 10^{-12}$ | $0.42 \times 10^{-11}$ | $0.17 \times 10^{-7}$ | $0.81 \times 10^{-7}$ | — | $0.15 \times 10^{-6}$ | $0.10 \times 10^{-1}$ |
| FL01 | $0.13 \times 10^{-14}$ | $0.41 \times 10^{-12}$ | $0.23 \times 10^{-9}$ | $0.92 \times 10^{-5}$ | $0.24$ | — | 0 |
| OP01 | 0 | 0 | 0 | 0 | 0 | 0 | — |

Table 6. K-S test of the X-ray contours

|      | SCDM | FL03 | OP03 | FL02 | OP02 | FL01 | OP01 |
|------|------|------|------|------|------|------|------|
| SCDM | — | $0.10 \times 10^{-8}$ | 0 | 0 | 0 | 0 | 0 |
| FL03 | $0.64 \times 10^{-1}$ | — | $0.21 \times 10^{-3}$ | $0.74 \times 10^{-3}$ | $0.30 \times 10^{-14}$ | $0.52 \times 10^{-4}$ | 0 |
| OP03 | $0.75 \times 10^{-6}$ | $0.39 \times 10^{-3}$ | — | $0.76$ | $0.52 \times 10^{-4}$ | $0.63$ | 0 |
| FL02 | $0.61 \times 10^{-4}$ | $0.14 \times 10^{-1}$ | $0.26$ | — | $0.47 \times 10^{-6}$ | $0.50$ | 0 |
| OP02 | 0 | $0.24 \times 10^{-13}$ | $0.92 \times 10^{-5}$ | $0.25 \times 10^{-8}$ | — | $0.92 \times 10^{-5}$ | $0.92 \times 10^{-5}$ |
| FL01 | 0 | 0 | $0.13 \times 10^{-7}$ | $0.41 \times 10^{-12}$ | $0.71 \times 10^{-1}$ | — | 0 |
| OP01 | 0 | 0 | 0 | 0 | $0.84 \times 10^{-13}$ | $0.70 \times 10^{-14}$ | — |